\documentclass[aps, prb, a4paper, reprint, 10pt, showpacs, twocolumn, groupedaddress, longbibliography]{revtex4-1}

\usepackage{amssymb}
\usepackage{amsmath}
\usepackage{amsfonts}
\usepackage{subcaption}
\usepackage{graphicx}
\usepackage{bm}
\usepackage{xcolor}
\usepackage{multirow}
\usepackage{natbib}
\usepackage{hyperref}
\usepackage{mathrsfs}
\usepackage{blkarray}
\usepackage{bbm}
\usepackage{tabularx}
\usepackage{dsfont, dcolumn, times, color, braket}

\newcolumntype{C}[1]{>{\centering\arraybackslash}m{#1}}
\begin{document}
\title{Spin-Weyl quantum unit: theoretical proposal}

\author{Y. Chen}
\affiliation{Kavli Institute of Nanoscience, Delft University of Technology, 2628 CJ Delft, The Netherlands}

\author{Y. V. Nazarov}
\affiliation{Kavli Institute of Nanoscience, Delft University of Technology, 2628 CJ Delft, The Netherlands}

\begin{abstract}
We propose a four-state quantum system, or quantum unit, that can be realized in superconducting hetero-structures. 
The unit combines the states of a spin and an Andreev qubit providing the opportunity of quantum superpositions of their states.  
This functionality is achieved by tunnel coupling between a 4-terminal superconducting heterostucture housing a Weyl point, and a quantum dot.
The quantum states in the vicinity of the Weyl point are extremely sensitive to small changes of superconducting phase, this gives reach opportunities for quantum manipulation.

We establish an effective Hamiltonian for the setup and describe the peculiarities of the resulting spectrum.  We concentrate on the 4-state subspace and explain how to make a double qubit in this setup.

We review various ways to achieve quantum manipulation in the unit, this includes resonant, adiabatic, diabatic manipulation and combinations of those. We provide detailed illustrations of designing arbitrary quantum gates in the unit.

\end{abstract}

\maketitle

\section{Introduction}
\label{sec:intro}
Superconducting qubits are defined in the micro-fabricated macroscopic-scale superconducting circuits with quantum properties. Such circuits generally comprise superconducting loops with weak link coupling the superconductors. The artificial quantum mechanics emerging from an interplay of Josephson effect and Coulomb blockade makes it possible  a rich variety of qubit designs. \cite{QuantumTransport}
Flux qubit\cite{PhysRevB.75.140515, PhysRevB.81.134510}, charge qubit\cite{PhysRevLett.101.080502, PhysRevB.76.174516} and phase qubit \cite{PhysRevB.77.214510, PhysRevB.67.100508} have been developed over the decades. The qubits defined in the circuits may be arranged to couple a common resonator mode, this enables multi-qubit quantum gates and non-invasive qubit measurements \cite{DiCarlo2009}.

Another major direction in solid-state quantum information processing are spin qubits, where the electon spin is used to store quantum information\cite{SpinQubitReview}. The spin qubits are usually realized in quantum dots in semiconductor materials where the electrons are confined in visibly discrete states. Both singlet\cite{PhysRevA.57.120, PhysRevLett.98.050502} and spin doublet\cite{PhysRevB.86.045316,PhysRevB.82.075403} schemes have been realized. The important experiments include \cite{nnano.2014.153, PhysRevB.96.165301, PhysRevLett.112.026801, PhysRevLett.118.167204}. The spin coherence time of these quantum dot systems may achieve milliseconds, which is beneficial for the quantum manipulation and quantum memory.


A less common but promising design of superconducting qubits exploits Andreev bound states: the localized quasiparticle states in the vicinity of superconducting contacts. It has been realized that with the Andreev bound states one can realize both kinds of the qubits within the same device. Namely, if the number of excess localized quasiparticles is even, a (an Andreev) qubit emerges from the ground and excited spin-singlet states \cite{zazunov:qubit}. However, if the number of excess quasiparticles is odd, the superconducting device houses a conveniently isolated spin qubit \cite{NazarovSpinQubit}. Such realization is more interesting than a traditional electron confinement in quantum dots motivating theoretical research \cite{PhysRevB.81.144519, EPL.100.57006}. These ideas have been realized experimentally \cite{Bretheau2013, janvier:science15, tosi:prx19} and remain in focus of attention of the superconducting qubit community.

Recently, a topological singularity in Andreev spectrum of multi-terminal superconducting structure --- a Weyl point --- has been predicted and theoretically investigated. \cite{WheylPointRiwar}. For a 4-terminal structure, the spectrum of Andreev states depends on three independent superconducting phases. At a particular choice of these three phases, the energy of the lowermost Andreev level approaches zero signalling the degeneracy of the corresponding spin-singlet qubit. The spectrum is conical in the vicinity of this singularity manifesting the critical dependence of the wave functions: very small changes of the phases in the vicinity of the point strongly affect the wavefunctions of the states. This is already advantageous for quantum manipulation applications. The Weyl points in the superconducting structures have been investigated in \cite{Weyl1,Weyl2,Weyl3,Weyl4,PhysRevB.99.165414}.

As for any Andreev-based setup, the parity effect is crucial here. For even parity, the spectrum of two spin-singlets is conical and the dependence of the wave functions is critical in the vicinity of the point. For odd parity, the spin-doublet states are slightly split owing to spin-orbit interaction.\cite{Yokoyama} Their wave functions or energies exhibit no critical dependence on the phases. (Fig. 1a)

\begin{figure}
\begin{center}
\includegraphics[width=\columnwidth]{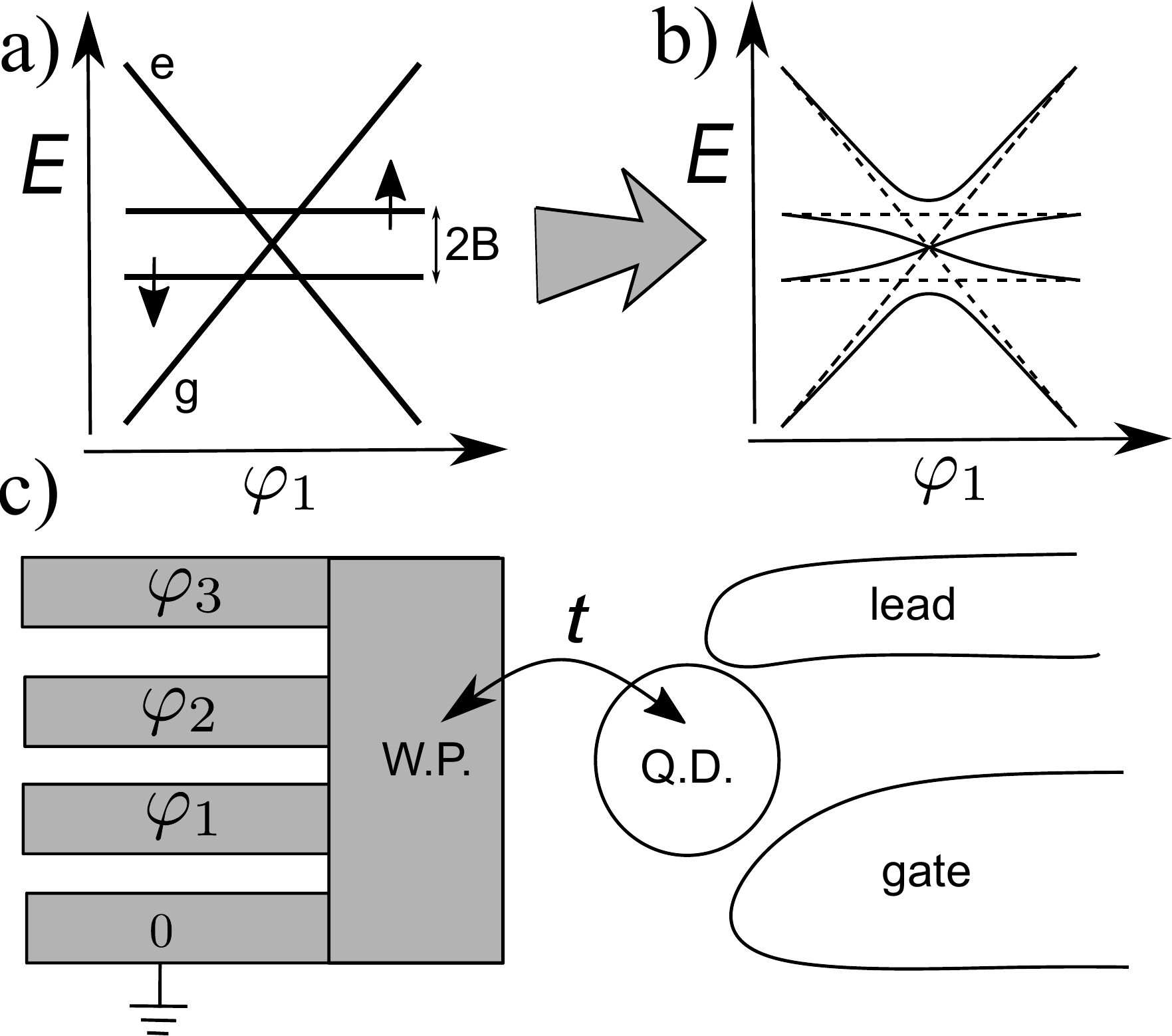}
\end{center}
\caption{The spin-Weyl quantum unit. a. The low-energy spectrum of the multi-terminal superconducting structure with a Weyl point consists of a pair of flat spin-doublet states an a pair of conical spin-singlets. The spin-doublets are split by small energy $2 B$ coming from spin-orbit interaction. The pairs are not coherent corresponding to different parities. b. The spin-Weyl quantum unit provides coherence and hybridization between the flat and conical states. Sketch of the spectrum. c. The setup of the spin-Weyl quantum unit. The superconducting structure with 4 leads and 3 independent phases $\varphi_{1,2,3}$ is tunnel-coupled with a single-electron quantum dot. The quantum dot is tuned by the gate electrode, a lead supplies electrons to the dot.}
\label{system}
\end{figure}%

The quantum spaces of different parity are completely separated and cannot be made coherent: indeed, a transition between those would involve a quasiparticle coming from/ escaping to the delocalized states of the continuous spectrum. So, despite the fact that the system can house both superconducting and spin qubit, there is no quantum coherence between the two.

Whatever tempting such coherence may be, it seems to be forbidden by fundamental laws.
The main point of this Article is that the coherence can be achieved with a rather simple extension of the Weyl point setup.

In this Article we propose a hybrid system that can be regarded as two coherently coupled qubits. It thus exhibits hybridization of flat spin-doublet states with conical spin-singlet states (Fig. 1 b.). We term the system a {\it spin-Weyl quantum unit}. We show how the unit can be manipulated to achieve an arbitrary unitary transformation in the space of 4 states, by the superconducting phase controls only, and can be conveniently read out.

The system proposed combines a superconducting heterostructure and a single-electron quantum dot (Fig.\ref{system} c). The two parts are coupled with a weak electron tunnelling between the heterostructure and the dot. The degeneracy at the Weyl point guarantees that even the weak coupling results in strong change of the spectrum making it advantageous for quantum manipulation applications. The electron tunneling to localized states of the dot is essential for breaking the parity conservation that forbids the hybridization of flat and conical states near the Weyl point.


The structure of the Article is as follows. In Section \ref{sec:Hamiltonian} we descibe the setup , establish a minimum Hamiltonian required to describe the unit, and explain its relevance in the wider context of more detailed description of the device. In Section \ref{sec:Spectrum}, we describe the resulting spectrum and the choice we made for the subspace where quantum manipulations are performed. We discuss in Section \ref{sec:Method} read-out, initialization, and various methods of quantum manipulation that can be implemented in the unit. 
In Section \ref{sec:sqgates}, we describe the implementation of single-qubit gates by means of resonant manipulation. 
In Section \ref{sec:2qgates} we concentrate on diabatic manipulations and demonstrate the design of various two-qubit gates. We conclude in Section \ref{sec:Conclusions}.

\section{The setup and the Hamiltonian}
\label{sec:Hamiltonian}
The unit consists of two subsystems: (Fig. \ref{system} c) the superconducting heterostructure  and the quantum dot. The superconducting heterostructure is connected to four superconducting leads biased with three independent superconducting phases $\varphi_{1,2,3}$ and houses the discrete Andreev bound states with the spectrum depending on the phases. At a certain choice of the phases, the energy lowermost Andreev state reaches zero exhibiting  a Weyl singularity \cite{WheylPointRiwar, PhysRevB.99.165414}. The quantum dot houses discrete number of electrons, this number can be tuned by a nearby gate electrode. A normal-metal lead supplies the electrons. The subsystems are coupled by electron tunneling.

Such setup can be realized with by a variety of technologies, including the 2D semiconducting heterostructures, semiconducting nanowires, graphine, brought in proximity with superconducting metals. Here we do not specify a concrete technology but rather spell out the physical restrictions. The spacing $\delta$ between the Andreev bound states becomes small in comparison with the superconducting energy gap $\Delta$ if the conductances $G$ in the superconducting structure exceed much the conductance quantum $G_Q$, $\delta \simeq (G_Q/G) \Delta$. So the conductances should be of the order of $G_Q$. As mentioned in \cite{WheylPointRiwar}, the existence of Weyl points relies on general scattering theory and does not impose any further restrictions on the properties of the structure. The vicinity of the Weyl point implies the working energy scale $\ll \Delta$. The level spacing in the quantum dot should be big at this energy scale. The tunneling energy should be also at this scale, that is, not too large. The spin-orbit splitting $B$ is much smaller than $\Delta$ in materials with weak spin-orbit coupling. The tunneling energy should be comparable with $B$.

Let us consider the full Hamiltonian of the system, that is the sum of the Weyl point structure Hamiltonian, the quantum dot Hamiltonian, and the tunneling Hamiltonian. 
\begin{equation}
H = H_{{\rm wp}} + H_{{\rm T}} + H_{{\rm QD}}
\end{equation}
We will construct a minimum Hamiltonian disregarding possible higher-energy states in the dot and in the structure. The Andreev levels in the vicinity of the Weyl point are described by a Weyl BdG Hamiltonian\cite{PhysRevB.99.165414}. Assuming spin degeneracy, this Hamiltonian is a $2\times 2$ matrix in Nambu space. Its general linear expansion near the Weyl point reads
\begin{equation}
\hat{H}^{{\rm WP}} = \hat{\tau}^{a} v_{ab} \delta \varphi_b
\end{equation} 
$a,b = 1,2,3$, $\hat{\tau}^{a}$ being a 3-vector of Pauli matrices in Nambu space, $\delta \varphi_b$ being small deviations of the superconducting phases from the Weyl point, $v_{ab}$ being a matrix of proportionality coefficients. It is advantageous to introduce the convenient coordinates in the space of 3 superconducting phases, $\phi^{a} = v_{ab} \delta \varphi_b$. We will name these coordinates {\it phases} for brevity, although they have dimension energy. The spectrum of the Hamiltonian is conveniently isotropic in the resulting space, $E = \pm |\vec{\phi}|$, while the wave functions do depend on the direction. To account for spin-orbit interaction, we promote the Hamiltonian to $4\times 4 $ matrix in spin and Nambu space,
\begin{equation}
\hat{H}^{{\rm WP}} = \vec{\hat \tau}\cdot \vec{\phi} + \frac{1}{2} \vec{B} \cdot \vec{\hat \sigma},
\end{equation} 
$\vec{\hat \sigma}$ being the vector of Pauli matrices in spin space.
The spin and orbital degrees of freedom separate, so the spectrum reads $E=\pm |\vec{\phi}| + s_z 2 B$, $B \equiv |\vec{B}|$, $s_z =\pm 1/2$ being the spin projection on the direction of $\vec{B}$. 
We need the Hamiltonian in second quantization form. We introduce the quasiparticle annihilation operators $\hat{\gamma}_\sigma$ and associated Nambu bispinors $\bar{\gamma}_{a,\sigma} \equiv (\hat{\gamma}_\sigma, \sigma \hat{\gamma}_{-\sigma})$ to recast it to the standart form,
\begin{equation}
H_{{\rm WP}} = \frac{1}{2} \bar{\gamma}^\dagger_\alpha \hat{H}^{{\rm WP}}_{\alpha \beta}  \bar{\gamma}_\beta
\end{equation}
This Hamiltonian can be reduced to a diagonal form for a certain direction in $\phi$-space, $\vec{\phi} = \phi {\vec {n}}$ by a Bogoliubov transform of $\hat{\gamma}_\sigma$ to a direction-dependent $\hat{\tilde{\gamma}}_\sigma$. Choosing the spin quantization axis along $\vec{B}$, we arrive at
\begin{equation}
H_{{\rm WP}} = \frac{1}{2}(\phi +B\sigma) \left(\hat{\tilde{\gamma}}_\sigma^\dagger \hat{\tilde{\gamma}}_\sigma - \hat{\tilde{\gamma}}_\sigma \hat{\tilde{\gamma}}^\dagger_\sigma\right)
\end{equation}
This gives the spectrum sketched in Fig. \ref{system} a: $E = \pm \phi$ for the states $|0\rangle$, $|2_{\uparrow\downarrow}\rangle \equiv \hat{\tilde{\gamma}}^\dagger_{\uparrow} \hat{\tilde{\gamma}}^\dagger_{\downarrow}|0\rangle$, $E = \pm B$ for the states $|\uparrow\rangle \equiv \hat{\tilde{\gamma}}^\dagger_{\uparrow}|0\rangle$, $|\downarrow\rangle  \equiv \hat{\tilde{\gamma}}^\dagger_{\downarrow}|0\rangle$. 

To model the dot at the small energy scale, we only need to take into account three states that differ by an addition of an electron: a non-degenerate state $|0\rangle$ and two spin-degenerate states $\hat{d}^\dag_\sigma|0\rangle$, $\hat{d}^\dag_\sigma$ being the electron creation operator corresponding to spin $\sigma$. The charging energy of the quantum dot pushes the states of other occupancy too high in energy. As such, the Hamiltonian reads
\begin{equation}
H_{{\rm QD}} = \epsilon_d \hat{d}^\dag_\sigma|0\rangle \langle 0| \hat{d}_\sigma,
\end{equation}
the value of $\epsilon_d$ can be tuned with the gate voltage. We disregard tunneling to the lead at the Hamiltonian level, although it is necessary for equilibration, in particular, for setting the overall parity in the setup.

The least trivial and the most important part of the total Hamiltonian describes tunnelling between the dot and the setup. To derive it, we assume spin conservation regarding spin-orbit as an irrelevant perturbation. Then the most general form of the tunneling Hamiltonian reads as follows:
\begin{equation}
H_{{\rm T}} = \int d {\bf y} d {\bf x} \hat{c}^\dagger_\sigma({\bf y}) \hat{d}({\bf x})_\sigma t({\bf x}, {\bf y}) + h.c. 
\end{equation}
Here, ${\bf y}$ and ${\bf x}$ are the electron coordinates in the superconducting structure and in the dot, respectively, $\hat{c}_\sigma({\bf y})$, $\hat{d}({\bf x})_\sigma$ are the corresponding electron annihilation operators, and $t({\bf x}, {\bf y})$ is the tunneling amplitude from the point ${\bf x}$ to the point ${\bf y}$. We need to project this operator to the low-energy electron states involved. To this end, we express the operators in terms of the wave functions of the quasiparticle states in the superconducting structure, and the electron states in the dot, and the corresponding creation/annihilation operators,
\begin{eqnarray}
\hat{c}_\sigma({\bf y})&=&\sum_n(u_n({\bf y})\hat{\gamma}_{n,\sigma}-\sigma v^*_n({\bf y})\hat{\gamma}^\dag_{n,-\sigma} );\\
\hat{d}_\sigma({\bf x}) &=&\sum_n \Psi_n({\bf x}) \hat{d}_n
\end{eqnarray}
where the summation is over all possible states. From this sum, we pick up the low-energy states, one for the superconducting structure, one for the dot, to arrive at:
\begin{equation}
H_{{\rm QD}} = \left(t_1 \hat{\gamma}^\dag_\sigma - t_2 \hat{\gamma}^\dag_{-\sigma}\right)\hat{d}_\sigma + h.c.
\end{equation}
with 
\begin{eqnarray}
t_1&=& \int d {\bf y} d {\bf x} u^*({\bf y}) \Psi({\bf x})  ;\\
t_2&=& \int d {\bf y} d {\bf x} v({\bf y}) \Psi({\bf x}) 
\end{eqnarray}
The tunnel Hamiltonian is thus characterized with two complex effective tunnelling amplitudes, $t_{1,2}$, whose common phase is irrelevant. It is important to understand that the remaining three parameters define the overall tunnelling strength $T \equiv \sqrt{|t_1|^2 +|t_2|^2}$ and the {\it direction} in the phase space. Thus, the tunneling breaks the isotropy in the phase space.

To analyse the spectrum, it is convenient to make the isotropy breaking explicit. For this, we fix the 3rd  axis of the coordinate system to the direction defined by $t_{1,2}$, express $\phi$ in spherical coordinates, $\vec{\phi} = \phi (\sin \theta \cos\mu, \cos \theta \sin \mu, \cos \theta)$, and perform the unitary transformation of $\hat{\gamma}_\sigma$ that diagonalizes $H_{{\rm WP}}$. With this, the transformed Hamiltonian reads
\begin{eqnarray}
H&= &\frac{1}{2}(\phi +B\sigma) \left(\hat{\tilde{\gamma}}_\sigma^\dagger \hat{\tilde{\gamma}}_\sigma - \hat{\tilde{\gamma}}_\sigma \hat{\tilde{\gamma}}^\dagger_\sigma\right) ] \\
&& +T \left( \cos\left(\frac{\theta}{2}\right) e^{i\mu/2} \hat{\tilde{\gamma}}^\dagger_\sigma \hat{d}_\sigma -\right. \notag \\
&& \left.\sin\left(\frac{\theta}{2}\right) e^{i\mu/2} \sigma\hat{\tilde{\gamma}}_{-\sigma} \hat{d}_\sigma+ h.c.\right) \notag \\
&+& H_{{\rm QD}}\notag
\end{eqnarray}
The azimuthal angle $\mu$ is not relevant for the spectrum and can be set to $0$. The spectrum depends on the polar angle $\theta$ owing to the tunnelling term.

\section{The spectrum}
\label{sec:Spectrum}
The whole spectrum consist of $3 \times 4 =12$ states. They are separated into two groups of different total parity, 6 states in each group. In addition, the states are separated by the spin projection $s_z$ on the B axis.

Let us consider the even parity first. There are two states with $s_z = \pm 1$, $|1_{\uparrow} 1_{\uparrow}\rangle$ and $|1_{\downarrow} 1_{\downarrow}\rangle$ (first and second number refer to the occupation of the superconducting structure and the dot, respectively) that are not affected by superconducting phases or tunnelling, with energies $\epsilon_d \pm B$. The group of four states with $s_z=0$ is of primary interest for us and comprises the spin-Weyl unit. Without tunnelling, there are two conical states, $|00 \rangle$, $|2_{\uparrow\downarrow} 0\rangle$  with energies $\pm \phi$, and two flat states $|1_{\uparrow} 1_{\downarrow}\rangle$ and $|1_{\downarrow} 1_{\uparrow}\rangle$ with energies $\epsilon_d \pm B$ (the second number in this notation is the occupation of the dot). The tunnelling hybridizes the states. The hybridization ceases at sufficiently large distances from the Weyl point, the energies of the states returning to their values without tunnelling. The Hamiltonian in this basis is given by the matrix (assuming $\mu =0$):
\begin{align}
H_4=\begin{bmatrix}
- \phi & -T\sin(\frac{\theta}{2}) &  T\sin(\frac{\theta}{2}) & 0\\
-T\sin(\frac{\theta}{2}) & \epsilon_{\rm d}-B & 0 &T\cos(\frac{\theta}{2}) \\
T\sin(\frac{\theta}{2})& 0 & \epsilon_{\rm d}+B & -T\cos(\frac{\theta}{2})\\
 0 & T\cos(\frac{\theta}{2}) & -T\cos(\frac{\theta}{2}) & \phi 
\label{parity-even}
\end{bmatrix}
\end{align}

\begin{figure*}
\includegraphics[width = 0.8\textwidth]{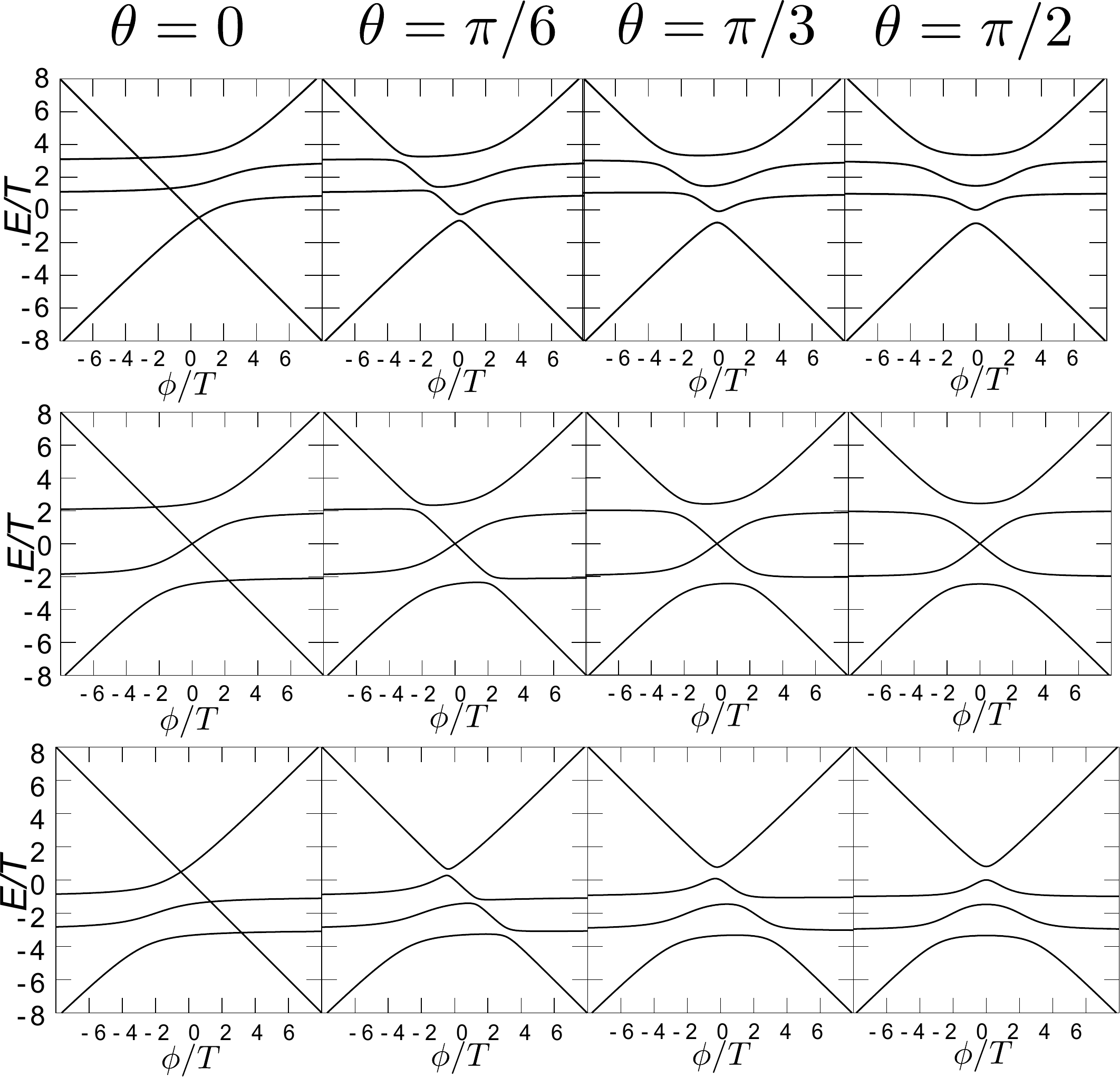}
\caption{The spectrum of spin-Weyl unit (even parity, $s_z=0$) consists of 4 sub-bands connected by three Weyl points, and emerges from hybridization of two flat and two conical subbands. The spectrum has axial symmetry. The columns correspond to four different settings of the polar angle $\theta$. The choices of parameters for rows: upper row, $B=T$, $\epsilon_d = 2T$, both flat bands are above the conical point; middle row, $B=2 T$, $\epsilon_d =0$, the conical point is between two flat subbands and remains at $\vec{\phi}=0$ for this parameter choice; lower row, $B=T$, $\epsilon_d = - 2T$, both flat bands are below the conical point.}
\label{fig:4spectrum}
\end{figure*}

The resulting axially symmetric spectrum is shown in Fig. \ref{fig:4spectrum} for various directions in the phase space given by the polar angle $\theta$ (negative $\phi$ correspond to polar angle $\pi -\theta$) and three different settings of $\epsilon_d$ where the former conical point is above, below or in between the energies of the flat states. 

The spectrum comprises 4 sub-bands that are eventually touch each other in 3 Weyl points. They are located at symmetry axis corresponding $\theta=0$ (or $\theta = \pi$, if $\phi <0$), leftmost column of the plots. For the middle row of the plots, the Weyl point is visible in all columns since it is located at $\vec{\phi}=0$ for a particular symmetric choice $\epsilon_d =0$ made. In general, the points are shifted from $\vec{\phi}=0$. The existence of these points is a consequence of topology, so these points remain even if we perturb the Hamiltonian, for instance, with the terms breaking the axial symmetry. Apart from this feature, the sub-bands show rather expected hybridization at $\phi \simeq T$ and go asymptotically to flat and conical states for  $\phi \gg T$. 

This 4-dimensional subspace suits well to represent two coupled qubits, and we will use it to realize the spin-Weyl quantum unit. 

For completeness, let us also describe the spectrum for odd parity. Six states are separated in two groups of three with $s_z = \pm 1/2$, that is, into two qutrits. The qutrit with $s_z =1/2$ is composed from the flat state $|1_{\uparrow} 0\rangle$, and the conical states $|0 1_\uparrow \rangle$, $|2_{{\uparrow}{\downarrow}} 1_\uparrow\rangle$ with energies $B$, $\epsilon_d \pm \phi$. The Hamiltonian is a $3 \times 3$ matrix:
\begin{align}
H_3 = \begin{bmatrix}
-\phi +\epsilon_d & -T \cos (\frac{\theta}{2}) & 0 \\
-T \cos (\frac{\theta}{2}) & B & - T \sin (\frac{\theta}{2})\\
0 & - T \sin (\frac{\theta}{2}) & \phi+\epsilon_d
\end{bmatrix}
\end{align}
\begin{figure*}
\includegraphics[width = 0.8\textwidth]{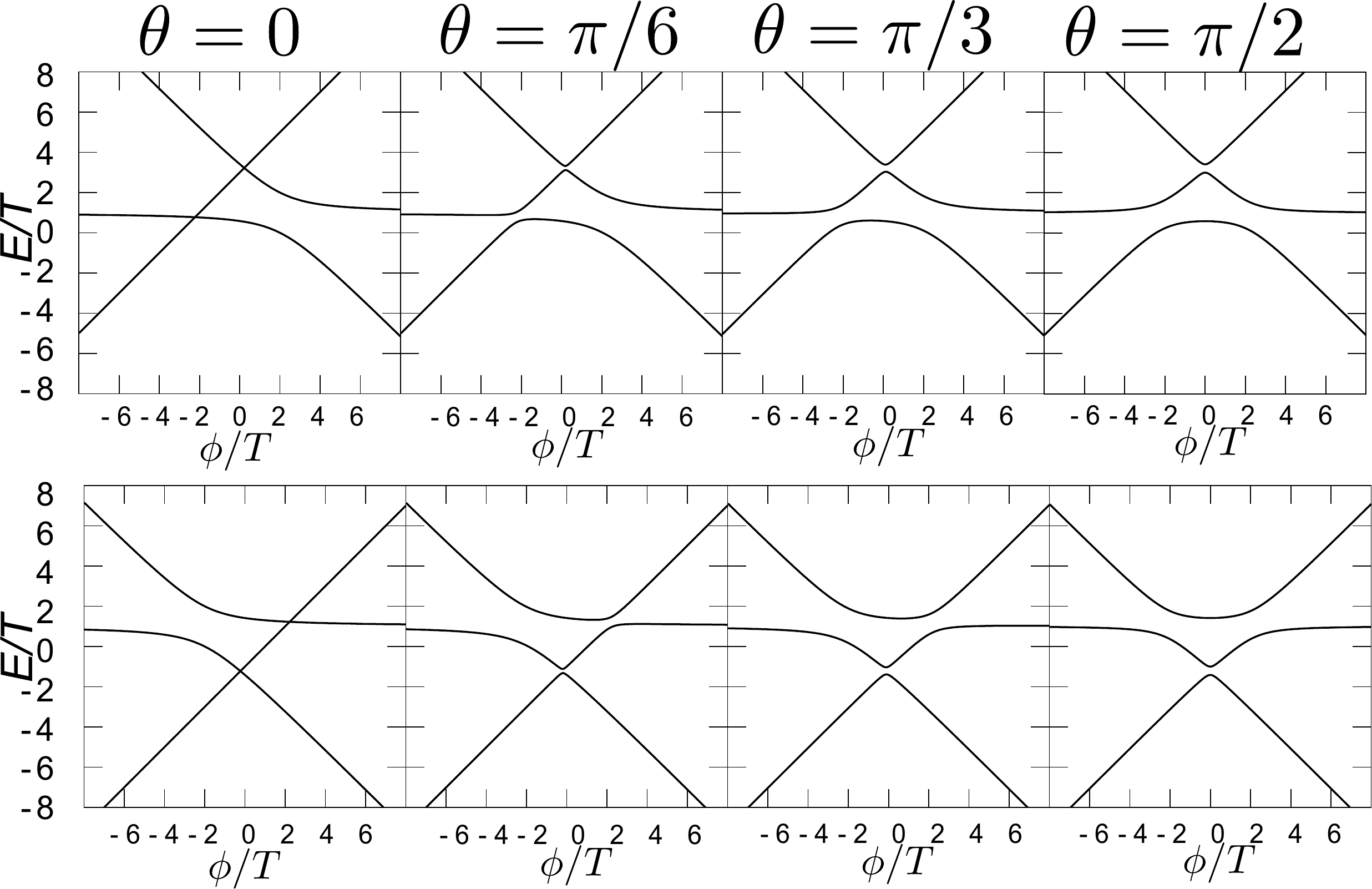}
\caption{The spectrum of a qutrit (odd parity, $s_z=1/2$) consists of 3 sub-bands connected by two Weyl points, and emerges from hybridization of one flat and two conical sub-bands. The spectrum has an axial symmetry. The columns correspond to four different settings of the polar angle $\theta$. The choices of parameters for rows: upper row, $B=T$, $\epsilon_d = 3T$, the flat band is above the conical point; lower row, $B=T$, $\epsilon_d =-T$, the flat band is  below the conical point.}
\label{fig:3spectrum}
\end{figure*}
The  spectrum is plotted in Fig. \ref{fig:4spectrum} for a set of polar angles and three different settings of $\epsilon_d$ where the former conical point is above or below the energy of the flat state.
There are two Weyl points in the spectrum that are situated at the axis. Apart from the number of flat subbands, the spectrum is similar to that of the spin-Weyl unit.
The spectrum of the qutrit with $s_z = -1/2$ is very similar to be obtained by inverting the value of $B$.




\section{Quantum information processing}
\label{sec:Method}
Let us discuss the system under consideration from the point of view of quantum information processing. Without going to unnecessary details, we describe all elements required for the processing: state read-out, initialization,  and various methods of manipulation.

{\it Read-out}. The most natural read-out in the system utilizes the supercurrents induced in the superconducting leads. For the state $i$, the supercurrent in the lead $j$ is given $I_j = \frac{2e}{\hbar} \frac{{\rm d}E_j}{{\rm d}\varphi_i}$. The supercurrents thus distinguish the slopes the states. A realistic measurement scheme is usual for superconducting qubits and involves a change of non-linear inductance in a resonator by this current, so the current is detected as a shift of resonant frequency \cite{janvier:science15}. One of the advantages of Weyl point is the conical property of the spectrum whereby the slops are of the same value far and close to the point, and can be significantly changed by a small change of the phase settings. The ground state and excited state in a conical pair give opposite supercurrent signals. In distinction, the spin-like flat states give almost zero supercurrent at $\phi \gg T$ and thus can not be distinguished. This is another advantage: the superposition of spin-like states is preserved by the measurement. Yet if necessary they can be distinguished as well: one needs to adiabatically change the phase settings close to zero where these states acquire slops owing to hybridization. 

{\it Initialization}. In the unit, one can adopt a conservative approach to initialization: just wait till the relaxation brings the system to the ground state. After this, one can go to the desired state by performing a manipulation. The problem may be that the relaxation without quasiparticle exchange in principle conserves parity, so the unit could stuck in the ground state of odd parity. In addition, the spin conservation in the process of relaxation may lead to stucking in $s_z =\pm 1$ states. To prevent this, one requires a tunnel connection to the lead which will change the parity and the spin projection. Another problem could be a slow relaxation from the flat states: this can be circumvented by setting the phases close to zero so these states are not flat any more.

{\it Manipulation}.
Let us see how we can manipulate the states in the unit. The most natural way is to change the superconducting phases in time. As mentioned, the advantage of Weyl point is that the big changes of the wave functions can be achieved by small $\phi \simeq T$ changes of phases. We do not consider manipulation by magnetic field that is typical for spin qubits since it is rather impractical: the magnetic fields required for such manipulation are much bigger than those required to provide the small superconducting phase changes. More interesting and practical possibility is to change the gate voltage modulating $\epsilon_d$, but we do not consider it here either.

\begin{figure}
\centering
\includegraphics[width=\columnwidth]{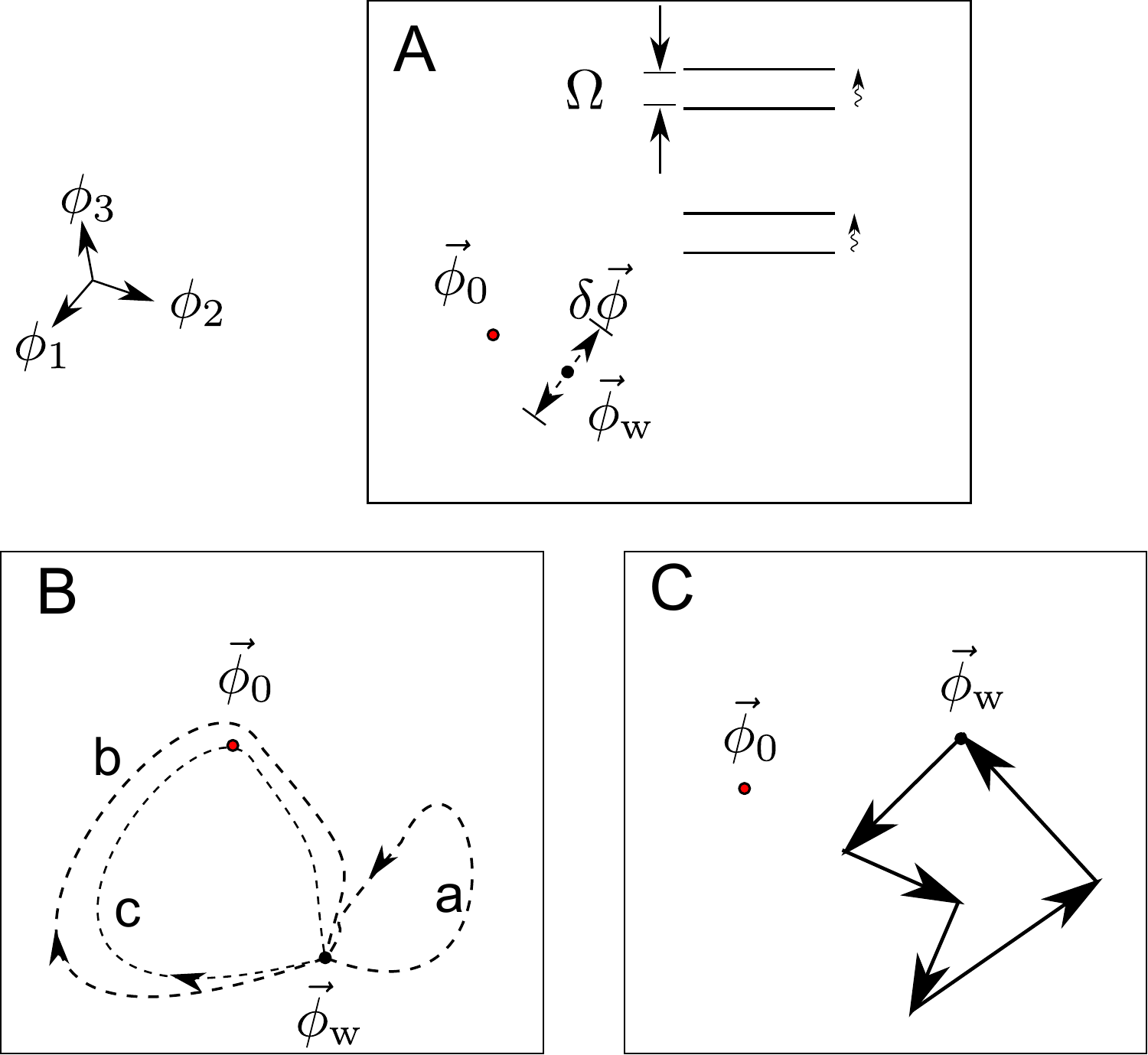}
\caption{The quantum manipulation methods for spin-Weyl quantum unit. A. Resonant manipulation. The phases oscillate around the working point $\phi_{{\rm w}}$ with amplitude $\delta \vec{\phi}$ and frequency $\omega$ that matches the level spacing $\Omega$. B. Adiabatic manipulation. Adiabatic change of phases along a trajectory can realize a phase gate (trajectory a) if a trajectory (trajectory a in the plot) is far from a Weyl point $\vec{\phi}_0$ or a Landau-Zener gate if a trajectory passes close to the point (trajectory b). Trajectory c passes directly through the point and realizes an exchange of wave function amplitudes in the sub-bands. C. Diabatic manipulation. The phase settings  jump from a working point $\vec{\phi}_{\rm w}$ to a series of consecutive points coming back in the end of the manipulation. }
\label{methods}
\end{figure}

The manipulation methods differ by the way the $\vec{\phi}$ is changing in time. Generally, there are three distinct methods: (Fig.\ref{methods}) i. resonant manipulation whereby a small oscillating  phase addition is applied at a working point $\vec{\phi}_{\rm w}$, ii. adiabatic manipulation whereby the phase is slowly changing along a trajectory in the phase space, usually returning to the initial point $\vec{\phi}$, iii. diabatic manipulation where the phases are changed by sudden jumps, and the changed settings are kept for a time interval before jumping back to another point or the initial point. Let us discuss each method for the unit in hand. We acknowledge that an efficient implementation of each method requires Hamiltonian characterization and subsequent design on the basis of a concrete Hamiltonian. However, the Hamiltonian is defined by a handful of parameters, so this should be a doable task. One can also employ the combinations of these methods.

{\it Resonant manipulation} is the most common manipulation method working for almost all quantum systems. If one applies a modulation $\delta \vec{\phi}(t)$ that oscillates with the frequency matching the energies of the states $|a\rangle$ and $|b\rangle$ defined in a working point $\phi_{\rm w}$, one is able to achieve an arbitrary unitary transformation in the basis $|a\rangle$, $|b\rangle$ tuning the duration and phase of the modulation pulse. \cite{QuantumTransport}
One needs to do more for a more general unitary transformation. To implement single-qubit gates, we use the resonant manipulation in special working points where two energy differences are the same, this permits unitary transformations in the basis of four states. We discuss the details in the Section \ref{sec:sqgates}.

{\it Adiabatic manipulation} involves a change of $\vec{\phi}$ along a closed trajectory. (Fig.\ref{methods}a) Usually, adiabaticity implies no transitions between the levels, this requires the velocity in phase space, $\vec{v} \equiv  \dot{\phi}$ to be small in comparison with the energy difference between the levels. For our system, where the energy difference in interesting region of the phase space are $\simeq T$, this implies $v \ll T^2$. With this, one can easily arrange a phase gate: an amplitude of a quantum state on each level acquires a phase factor with no change of its modulus (trajectory $a$ in Fig.\ref{methods}b). Notably, the presence of Weyl points in the spectrum of the unit permits design of more complicated gates. The point is that the level splitting becomes small near the point and, for any fixed $v$, the adiabaticity criterion is not satisfied if the trajectory comes sufficiently close to the Weyl point. (trajectory $b$ in Fig.\ref{methods}c) This realizes a Landau-Zener gate \cite{Oliver2005}: there is a non-adiabatic transition between two subbands with an amplitude $\alpha$ given by Landau-Zener formula, $|\alpha| = \exp( - \pi \phi^2_d/v)$, $\phi_d$ being the smallest distance between the Weyl point and the trajectory. If the trajectory goes precisely throught the Weyl point, a SWAP gate for two subbands is realized (trajectory $c$ in Fig.\ref{methods}b). The phase factors accumulated in the course of the Landau-Zener transition can be adjusted by tuning the shape or velocity at the returning trajectory.

{\it Diabatic manipulation} is implemented as a sequence of sudden jumps between the points in the phase space, that brings the system back to the initial point. (Fig.\ref{methods}c) The wave function does not change during the jumps. After each jump, the phase settings are kept constant during a time interval to let the wave function evolve with a Hamiltonian local to the point. To prevent the excitation to higher states in the course of jump, its actual duration should be yet longer than the inverse energy distance to higher levels. The Weyl point structure is advantageous for diabatic manipulation since big changes of the Hamiltonian can be produced by small jumps in the phase space.

This variety of manipulation methods permits multiple implementations of quantum gates. To illustrate, we have to make choices. 

To start with, we define a bipartition of 4-dimensional Hilbert space into two qubits. We label the states as $\left | 00\right\rangle$, $\left |01 \right\rangle$, $\left |10 \right\rangle$, $\left | 11\right\rangle$ from the lowest energy to the highest energy, that is, the second qubit has the smaller excitation energy $E_{01} - E_{00}$.  For separate non-interacting qubits, one expects $E_{11}+E_{00}=E_{01}+E_{01}$. This condition is generally not fulfilled for the spectrum in hand. However, it is fulfilled asymptotically at $\epsilon_d =0$ and also in special points of the phase space.

We assume $|\epsilon_d| <B$, so that the conical point is between the energies of the flat states. 

For the implementation of single-qubit gates, we choose resonant manipulation in special points. For the implementation of the two-qubit gates, we choose diabatic manipulation. We describe the implementations in the subsequent sections.

\section{The single-qubit Gates}
\label{sec:sqgates}

We will realize the single-qubit gates by means of resonant manipulation. This realization requires some tuning. Generally, the three energy differences between the levels are all different and resonant manipulation would result in a two-qubit gate. For instance, if the frequency of the oscillating field is in resonance with the energy difference between the states $|10\rangle$ and $|11\rangle$, one can swap by a pulse the amplitudes of the states $|10\rangle$ and $|11\rangle$ realizing the traditional CNOT gate \cite{QuantumComputation2000}.

The single-qubit manipulation is possible at a family of special working points where $E_{11}+E_{00}=E_{01}+E_{01}$, this corresponds to independent qubits with energy splittings $\Omega_1 = E_{10} - E_{00}$ and $\Omega_2 = E_{01} - E_{00}$. At these frequencies, the oscillating field resonates with two pairs of levels.  Since the energy spectrum is independent of the azimuthal angle, these special working points form a surface of revolution around $z$-axis.

Let us explain how one realizes the $X$ rotations of the first qubit. As an example, we take $B=3T$, $\epsilon_d = 2T$. The special working point can be realized at $\vec{\phi}_{{\bf w}}/T =(1.3,0,2.25)$. The qubit splittings are: 
$\Omega_1 =6.0 T$, $\Omega_2 =2.0 T$. 
We apply an oscillatory modulation $\delta\vec{\phi}(t) = {\rm Re}( \delta\vec{\phi} e^{i\Omega_1 t})$, $\delta \vec{\phi}$ being a complex vector of the oscillation amplitudes. It results in a time-dependent perturbation $\hat{h} = \frac{1}{2}\delta \vec{\phi}\cdot \vec{\tau} e^{i\Omega_1 t} + h.c.$. For the result of the manipulation not  to depend on the state of the second qubit, this perturbation should satisfy $\langle 10|\hat{h}|00 \rangle = \langle 11|\hat{h}|01 \rangle$. Since both matrix elements can be presented as scalar products of complex vectors, 
$h_{10,00} = \vec{v}_{10,00} \cdot \delta \vec{\phi}$ , and similar for another matrix elements.
To satisfy independence from the second qubit, the direction of $\delta \phi$ should be orthogonal to 
$\vec{v}_{10,00}-\vec{v}_{11,01}$. It also has to be orthogonal to the cross-product of the vectors, since the modulation in this direction does not appear in the matrix elements. For the example in hand, this fixes $\delta{\phi}$ to $|\delta \vec{\phi}| (0.16, 0.21 i , 0.97)$. To perform the rotation $\exp(i\gamma \sigma_x)$, one chooses 
$\gamma = 0.34 |\delta \vec{\phi}| {\cal T} $, ${\cal T}$ being the pulse duration.

To design the X-rotation of the second qubit, we proceed in the same way choosing the direction of osclillations to achieve $\langle 01|\hat{h}|00 \rangle = \langle 11|\hat{h}|10 \rangle$. 
This fixes $\delta{\phi}$ to $|\delta \vec{\phi}| (0.6, 0.7 i , -0.35)$. To perform the X-rotation $\exp(i\gamma \sigma_x)$, one chooses 
$\gamma = 0.04 |\delta \vec{\phi}| \tau $, $\tau$ being the pulse duration. 

As it is usual in the context of resonant manipulation, the Y and Z rotations can be achieved by changing the total phase of the oscillation and frequency modulation, respectively.

\section{The two-qubit Gates}
\label{sec:2qgates}
More complex gates require realization of an arbitrary $4\times4$ s-unitary transformations. In principle, this can be achieved only by means of resonance manipulation and adiabatic manipulation. However, this requires a tedious design and the time of the manipulation should greatly exceed the inverse energy differences $\simeq T^{-1}$. So we turn to diabatic manipulation. 

For diabatic manipulation, it is proficient to work with the spin-Weyl Hamiltonian in the phase-independent basis where it takes the form
\begin{align}
H_4=\begin{bmatrix}
- \phi_3 & 0 &  0 & \phi_1 +i \phi_2\\
0 & \epsilon_{\rm d}-B & 0 &T \\
0& 0 & \epsilon_{\rm d}+B & -T\\
 \phi_1-i\phi_2 & T & -T & \phi_3 
\end{bmatrix}
\end{align}
In this basis, the wave function remains continuous upon a diabatic change of $\vec{\phi}$.

The manipulation starts in a working point $\vec{\phi}_{\rm w}$ where the Hamiltonian is diagonalized as
\begin{equation}
H_4(\vec{\phi}_{\rm w}) = D E_d D^{-1}
\end{equation} 
$E_d$ being the diagonal matrix of two qubit eigenstates.
The phase then goes through a set of points $\vec{\phi}_i$ staying for a time interval $t_i$ in each point and finally returning to 

The result of the manipulation is a unitary $4 \times 4$ matrix in the basis of two-qubit eigenstates,
\begin{align}
S &= D^{-1} e^{i H_4(\vec{\phi}_{\rm w}) \sum_i t_i}\prod_i S_i e^{-i H_4(\vec{\phi}_{\rm w}) \sum_i t_i} D;\\
S_i &\equiv \exp(-i H_4(\vec{\phi}_i) t_i)
\end{align}

To design a manipulation given a target $S$, we need to choose 
$\vec{\phi}_i, t_i$ in a proper way. An arbitrary SU(4) transformation depends on $4^2-1=15$ parameters, while each jumping point brings 4 parameters: 3 phases and 1 time interval. Consequently, an arbitrary SU(4) transformation requres at least 4 jumping points(Fig. \ref{methods}C). 
 To accomplish the design task numerically, we specify the target unitary matrix $S_{\rm t}$ and define a minimization function in the space of the manipulation parameters $\{\vec{\phi}_i,t_i\}$,
\begin{equation}
U(\{\vec{\phi}_i,t_i\}) =8-{\rm Tr}(S_t S^\dag+S S_t^\dag) .
\end{equation}
We start the minimization routine with a random point in 16-dimensional space, iterate to a minimum and check if $U=0$ in this minimum. We accomplish this by setting a threshold of $U_{\rm th} \ll T^2$ and see if $U$ falls into the interval of $[
0,U_{\rm th}]$. If $0\leq U \leq U_{\rm th}$, we have found the solution: the minimum $U = 0$ is achieved only if $S=S_{t}$. If otherwise $U>U_{\rm th}$, we repeat the procedure starting another random point.%
 
A set of universal quantum gates can be achieved combining elementary quantum logic gates. The minimum circuit requirement for a general two-qubit manipulation can be constructed with 3 CNOT ($cX$) gates and 15 elementary one-qubit gates\cite{PhysRevA.69.032315}. 
In principle, the single-qubit gates can be also designed by diabatic manipulation method. However, we have already achieved these gates as described in the previous Section.  
 Here, we present the design of 3 controlled Pauli gates, $cX$, $cY$ and $cZ$ that $c(\sigma_i)=\begin{pmatrix}
\sigma_i&\\
&\mathbbm{1}
\end{pmatrix}$ in the qubit basis with the first qubit serving as control one.%

\begin{figure*} 
\begin{minipage}{0.36\linewidth}
\centering
\begin{tabular}{| l | p{4cm} | p{1.5cm} |}
  \hline	
Gate&Phase ($\phi/T$)& Time (t/T) \\
  \hline \hline
   $cX$ & $\phantom{\rightarrow}$ $\vec{\phi}_{\rm w}$ \newline $\rightarrow (1.04, -0.11, 1.07)$ \newline $\rightarrow(4.56, 6.76, 5.52)$ \newline $\rightarrow (6.62,  2.67, -1.11)$\newline $\rightarrow (5.24 ,  2.13,  6.34) $\newline $\rightarrow \vec{\phi}_{\rm w}$ & $\phantom{,}$\newline $\phantom{\rightarrow}$ 5.10 \newline $\rightarrow$ 5.96 \newline $\rightarrow$ 0.03 \newline $\rightarrow$ 7.85  \\
  \hline
  $cY$ & $\phantom{\rightarrow}$ $\vec{\phi}_{\rm w}$ \newline $\rightarrow(4.40, 9.82, 12.15)$ \newline $\rightarrow(5.63, 1.35, -0.87)$ \newline $\rightarrow(0.84 , -1.15, -0.46)$ \newline $\rightarrow(11.99, 14.25  ,  7.60)$ \newline $\rightarrow\vec{\phi}_{\rm w}$& $\phantom{,}$\newline $\phantom{\rightarrow}$ 2.80 \newline $\rightarrow 3.77$\newline $\rightarrow5.43$ \newline $\rightarrow2.05$ \\
  \hline
 $cZ$ & $\phantom{\rightarrow}$ $\vec{\phi}_{\rm w}$ \newline $\rightarrow(9.41, 2.82, 4.57)$ \newline $\rightarrow(-1.28, 0.18, 0.90)$ \newline $\rightarrow(9.72,  9.90, 10.22)$ \newline $\rightarrow(1.55,  0.91,  7.67)$ \newline $\rightarrow\vec{\phi}_{\rm w}$& $\phantom{,}$\newline $\phantom{\rightarrow}$ 7.18 \newline $\rightarrow6.96$ \newline $\rightarrow2.53$ \newline $\rightarrow5.27$ \\
  \hline
\end{tabular}
\end{minipage}%
\qquad%
\begin{minipage}{0.49\linewidth}
\centering
	\includegraphics[width=\textwidth]{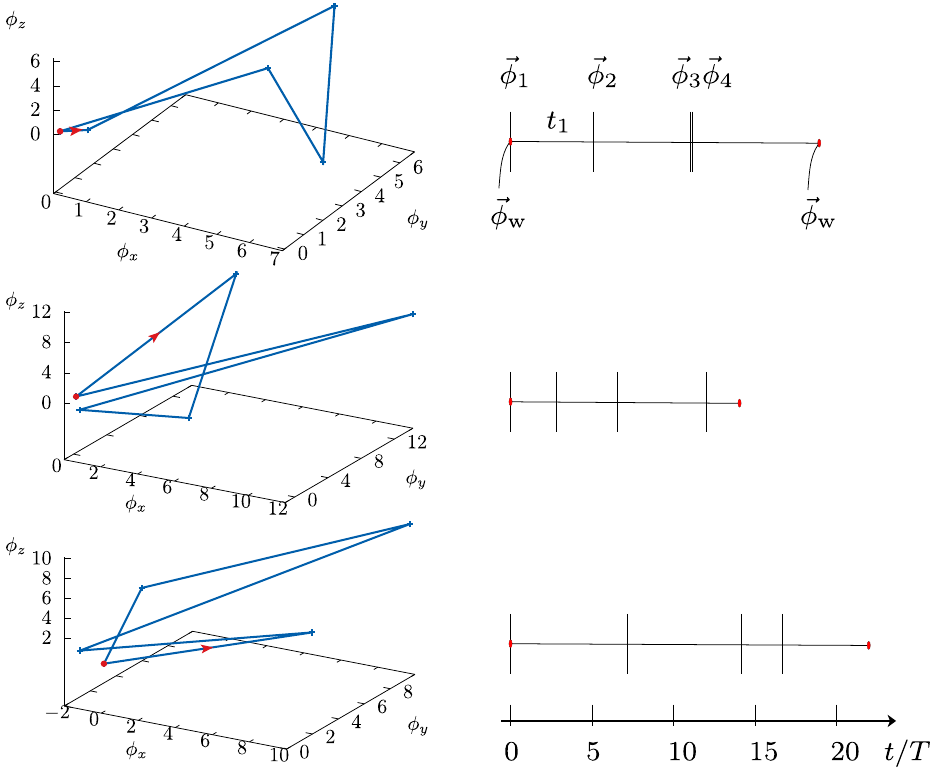}
\end{minipage}%
\caption{\label{diabatic}Design of two-qubit gates by diabatic manipulation. We implement controlled logic gates $cX$, $cY$ and $cZ$ gates, the first qubit being control one. The choice of the parameters: $B = 3T, \epsilon_d=2T$, working point: $\vec{\phi}_{\rm w} =(1.3,0,2.25)T$. The table specifies for each three gates the set of jumping points $\vec{\phi}_i$ and the time intervals $t_i$. The plots illustrate the diabatic paths and time intervals. The red mark in each graph indicates the initial working point. The arrow indicates the diabatic jumping sequence initiated from the working point.}
\end{figure*}

We choose the same parameters and the working point as in the previous Section: $B=3T$, $\epsilon_d = 2T$, $\vec{\phi}_{{\bf w}}/T =(1.3,0,2.25)$.
The results are presented in Fig. \ref{diabatic}. The advantage of the diabatic manipulation is the speed: the longest manipulation takes no more than $\simeq 20\,T^{-1}$.

\section{Conclusions}
\label{sec:Conclusions}
In conclusion, we propose a spin-Weyl quantum unit: a four-state system that can be regarded as a coherent combination of spin and Andreev superconducting qubits. The coherence, that seemingly breaks the parity conservantion, can be achived by coupling a 4-terminal superconducting structure housing a Weyl point to a quantum dot. We derive a simple but non-trivial universal Hamiltonian for the setup and choose 4-dimensional subspace for the realization of the spin-Weyl quantum unit. We have described the methods and adtantages of the quantum manipulation by controlling the superconducting phases in the vicinity of the Weyl point. We illustrate this by providing concrete designs of single-qubit and two-qubit quantum gates. 

Such devices can be fabricated and tuned, and, as it is common in superconducting qubit technologies, can be made work together in a many-unit quantum computer by coupling them to electric resonant modes. The system described calls for an experimental realization.

\begin{acknowledgements}
This project has received funding from the European Research Council (ERC) under the European Union's Horizon 2020 research and innovation programme (grant agreement \# 694272).
\end{acknowledgements}

\bibliographystyle{apsrev4-1}
\bibliography{spin-weyl-unit}
\end{document}